\documentclass[twocolumn,prd,superscriptaddress,altaffilletter,nofootinbib]{revtex4-1}
\usepackage{amsmath}
\usepackage{amssymb}
\usepackage{amsfonts}
\usepackage{comment}
\usepackage{graphicx}
\usepackage{hyperref}
\usepackage[nolist]{acronym}
\usepackage[usenames,dvipsnames]{xcolor}
\usepackage{xspace}
\usepackage{tikz}
\usepackage{lineno,xcolor}
\usepackage[squaren]{SIunits}
\usepackage{textcomp}
\usepackage[caption=false]{subfig}
\usepackage[export]{adjustbox}
\usetikzlibrary{arrows,shapes,trees,decorations.pathreplacing}
\allowdisplaybreaks

\newcommand{\likehood}{\mathcal{L}}

\newcommand{\params}{\bar{\theta}}
\newcommand{\sh}{\mathrm{signal}} 
\newcommand{\nh}{\mathrm{noise}} 

\newcommand{\pipeline}{GstLAL-based inspiral pipeline}

\newcommand{\msun}{\ensuremath{\mathrm{M}_{\odot}}\xspace}
\newcommand{\Mc}{\ensuremath{\mathcal{M}}}

\newcommand{\abs}[1]{\left\lvert#1\right\rvert}

\begin{document}

\title{The GstLAL Search Analysis Methods for Compact Binary Mergers in Advanced LIGO's Second and Advanced Virgo's First Observing Runs}

\author{Surabhi Sachdev}
\affiliation{Department of Physics, The Pennsylvania State University, University Park, PA 16802, USA}
\affiliation{Institute for Gravitation and the Cosmos, The Pennsylvania State University, University Park, PA 16802, USA}
\affiliation{LIGO Laboratory, California Institute of Technology, MS 100-36, Pasadena, California 91125, USA}

\author{Sarah Caudill}
\affiliation{Leonard E.\ Parker Center for Gravitation, Cosmology, and Astrophysics, University of Wisconsin-Milwaukee, Milwaukee, WI 53201, USA}
\affiliation{Nikhef, Science Park, 1098 XG Amsterdam, Netherlands}

\author{Heather Fong}
\affiliation{Canadian Institute for Theoretical Astrophysics, 60 St. George Street, University of Toronto, Toronto, Ontario, M5S 3H8, Canada}
\affiliation{RESCEU, The University of Tokyo, Tokyo, 113-0033, Japan}

\author{Rico K. L. Lo}
\affiliation{Department of Physics, The Chinese University of Hong Kong, Shatin, New Territories, Hong Kong}
\affiliation{LIGO Laboratory, California Institute of Technology, MS 100-36, Pasadena, California 91125, USA}

\author{Cody Messick}
\affiliation{Department of Physics, The Pennsylvania State University, University Park, PA 16802, USA}
\affiliation{Institute for Gravitation and the Cosmos, The Pennsylvania State University, University Park, PA 16802, USA}

\author{Debnandini Mukherjee}
\affiliation{Leonard E.\ Parker Center for Gravitation, Cosmology, and Astrophysics, University of Wisconsin-Milwaukee, Milwaukee, WI 53201, USA}

\author{Ryan Magee}
\affiliation{Department of Physics, The Pennsylvania State University, University Park, PA 16802, USA}
\affiliation{Institute for Gravitation and the Cosmos, The Pennsylvania State University, University Park, PA 16802, USA}

\author{Leo Tsukada}
\affiliation{RESCEU, The University of Tokyo, Tokyo, 113-0033, Japan}
\affiliation{Department of Physics, Graduate School of Science, The University of Tokyo, Tokyo, 113-0033, Japan}

\author{Kent Blackburn}
\affiliation{LIGO Laboratory, California Institute of Technology, MS 100-36, Pasadena, California 91125, USA}

\author{Patrick Brady}
\affiliation{Leonard E.\ Parker Center for Gravitation, Cosmology, and Astrophysics, University of Wisconsin-Milwaukee, Milwaukee, WI 53201, USA}

\author{Patrick Brockill}
\affiliation{Leonard E.\ Parker Center for Gravitation, Cosmology, and Astrophysics, University of Wisconsin-Milwaukee, Milwaukee, WI 53201, USA}

\author{Kipp Cannon}
\affiliation{Canadian Institute for Theoretical Astrophysics, 60 St. George Street, University of Toronto, Toronto, Ontario, M5S 3H8, Canada}
\affiliation{RESCEU, The University of Tokyo, Tokyo, 113-0033, Japan}


\author{Sydney J. Chamberlin}
\affiliation{Department of Physics, The Pennsylvania State University, University Park, PA 16802, USA}
\affiliation{Institute for Gravitation and the Cosmos, The Pennsylvania State University, University Park, PA 16802, USA}

\author{Deep Chatterjee}
\affiliation{Leonard E.\ Parker Center for Gravitation, Cosmology, and Astrophysics, University of Wisconsin-Milwaukee, Milwaukee, WI 53201, USA}

\author{Jolien D. E. Creighton}
\affiliation{Leonard E.\ Parker Center for Gravitation, Cosmology, and Astrophysics, University of Wisconsin-Milwaukee, Milwaukee, WI 53201, USA}


\author{Patrick Godwin}
\affiliation{Department of Physics, The Pennsylvania State University, University Park, PA 16802, USA}
\affiliation{Institute for Gravitation and the Cosmos, The Pennsylvania State University, University Park, PA 16802, USA}

\author{Anuradha Gupta}
\affiliation{Department of Physics, The Pennsylvania State University, University Park, PA 16802, USA}
\affiliation{Institute for Gravitation and the Cosmos, The Pennsylvania State University, University Park, PA 16802, USA}

\author{Chad Hanna}
\affiliation{Department of Physics, The Pennsylvania State University, University Park, PA 16802, USA}
\affiliation{Department of Astronomy and Astrophysics, The Pennsylvania State University, University Park, PA 16802, USA}
\affiliation{Institute for Gravitation and the Cosmos, The Pennsylvania State University, University Park, PA 16802, USA}
\affiliation{Institute for CyberScience, The Pennsylvania State University, University Park, PA 16802, USA}

\author{Shasvath Kapadia}
\affiliation{Leonard E.\ Parker Center for Gravitation, Cosmology, and Astrophysics, University of Wisconsin-Milwaukee, Milwaukee, WI 53201, USA}


\author{Ryan N.\ Lang}
\affiliation{Leonard E.\ Parker Center for Gravitation, Cosmology, and Astrophysics, University of Wisconsin-Milwaukee, Milwaukee, WI 53201, USA}

\author{Tjonnie G. F. Li}
\affiliation{Department of Physics, The Chinese University of Hong Kong, Shatin, New Territories, Hong Kong}

\author{Duncan Meacher}
\affiliation{Department of Physics, The Pennsylvania State University, University Park, PA 16802, USA}
\affiliation{Institute for Gravitation and the Cosmos, The Pennsylvania State University, University Park, PA 16802, USA}
\affiliation{Leonard E.\ Parker Center for Gravitation, Cosmology, and Astrophysics, University of Wisconsin-Milwaukee, Milwaukee, WI 53201, USA}



\author{Alexander Pace}
\affiliation{Department of Physics, The Pennsylvania State University, University Park, PA 16802, USA}
\affiliation{Institute for Gravitation and the Cosmos, The Pennsylvania State University, University Park, PA 16802, USA}

\author{Stephen Privitera}
\affiliation{Albert-Einstein-Institut, Max-Planck-Institut f{\"u}r Gravitationsphysik, D-14476 Potsdam-Golm, Germany}


\author{Laleh Sadeghian}
\affiliation{Leonard E.\ Parker Center for Gravitation, Cosmology, and Astrophysics, University of Wisconsin-Milwaukee, Milwaukee, WI 53201, USA}



\author{Leslie Wade}
\affiliation{Department of Physics, Hayes Hall, Kenyon College, Gambier, Ohio 43022, USA}

\author{Madeline Wade}
\affiliation{Department of Physics, Hayes Hall, Kenyon College, Gambier, Ohio 43022, USA}

\author{Alan Weinstein}
\affiliation{LIGO Laboratory, California Institute of Technology, MS 100-36, Pasadena, California 91125, USA}

\author{Sophia Liting Xiao}
\affiliation{LIGO Laboratory, California Institute of Technology, MS 100-36, Pasadena, California 91125, USA}

\date{\today}
\begin{abstract}
After their successful first observing run (September 12, 2015 - January 12,
2016), the Advanced LIGO detectors were upgraded to increase their sensitivity
for the second observing run (November 30, 2016 - August 26, 2017). The
Advanced Virgo detector joined the second observing run on August 1, 2017. We
discuss the updates that happened during this period in the \pipeline{}, which
is used to detect gravitational waves from the coalescence of compact binaries both
in low latency and an offline configuration. These updates include  deployment
of a zero-latency whitening filter to reduce the over-all latency of the
pipeline by up to 32 seconds, incorporation of the Virgo data stream in the
analysis, introduction of a single-detector search to analyze data from the
periods when only one of the detectors is running, addition of new parameters
to the likelihood ratio ranking statistic, increase in the parameter space of
the search, and introduction of a template mass-dependent glitch-excision
thresholding method.
\end{abstract}

\maketitle

\section{Introduction}\label{sec:intro}

The observations of the binary neutron star merger GW170817~\cite{PhysRevLett.119.161101} and the associated
short gamma ray burst GRB 170817A by the LIGO-Virgo Scientific Collaboration
and the Fermi-GBM monitor and INTEGRAL satellite~\cite{Monitor:2017mdv} led to an electromagnetic follow-up
on an unprecedented scale~\cite{2041-8205-848-2-L12} and marked the dawn of a new era of multi-messenger
astronomy.

The gravitational-wave event was identified
in low-latency by the \pipeline{}~\cite{messick2017analysis, cannon2012toward,
privitera2014improving}. The low-latency detection of the gravitational-wave
event  made it possible for alerts to be sent out to the electromagnetic
facilities in the timely fashion required for the identification of the optical
transient. These gravitational-wave and electromagnetic observations support
the hypothesis that neutron star mergers are progenitors of short Gamma-Ray
Bursts and are followed by transient electromagnetic events known as
kilonovae~\cite{2041-8205-848-2-L12}. The joint observations allow us to set
strong constraints on the fundamental physics of gravity~\cite{Monitor:2017mdv}, and also provide a
new method of probing the cosmological parameters~\cite{Abbott:2017xzu}.
Besides being vital for multi-messenger astronomy, low-latency detection is
also useful to freeze the detector state in case of a candidate, so as to collect
sufficient data for background estimation before any configuration changes are
made in the detectors. 

The \pipeline{} (henceforth referred to as the GstLAL pipeline) is a matched-filtering
analysis pipeline that can detect gravitational waves from compact binary
mergers in near real time, and provide point estimates for binary parameters.
Matched-filtering is performed by cross-correlating data against a bank of
waveform templates formed using general relativity. The GstLAL pipeline is built on
the GstLAL library, a collection of GStreamer~\cite{gstreamer} libraries and
plug-ins that make use of the LIGO Algorithm Library, LALSuite~\cite{lalsuite}.
It uses the GStreamer library to stream the gravitational strain data in real
time, performs matched-filtering in the time-domain~\cite{cannon2012toward,
messick2017analysis}, as opposed to the more traditional
frequency-domain~\cite{allen2012findchirp} method, and uses a time-domain
rather than a frequency-domain signal consistency
test~\cite{allen2012findchirp}. It also employs multi-banding and singular
value decomposition on signal templates~\cite{cannon2010singular,
messick2017analysis} to reduce the number of filters and samples used in
matched-filtering. A multidimensional likelihood-ratio statistic is used to
rank the gravitational-wave candidates according to the properties of noise and
signal~\cite{cannon2015likelihood, messick2017analysis}. Instead of performing
time-slides~\cite{allen2012findchirp} for background estimation, a technique
that is based on tracking the noise distributions and allows for rapid
significance estimation~\cite{cannon2013method} is used. Several other low-latency gravitational-wave pipelines
are used by the LIGO-Virgo Collaboration to analyze their data, MBTA Online~\cite{0264-9381-33-17-175012},
GstLAL-spiir~\cite{2017arXiv170202256G}, PyCBC Live~\cite{Nitz:2018rgo}, and coherentWaveBurst~\cite{PhysRevD.93.042004}.

In this work, we describe the \pipeline{}, which was used to detect the binary
neutron star merger event GW170817~\cite{PhysRevLett.119.161101} and several binary black hole mergers~\cite{Abbott:2017vtc, Abbott2017b, PhysRevLett.119.141101, LIGOScientific:2018mvr} during
the second observing run of the Advanced LIGO detectors and the first observing
run of the Advanced Virgo detector (henceforth referred to as the second observing run or simply, O2).
We focus on the updates made in the GstLAL pipeline since the first observing run of the
LIGO detectors (O1)~\cite{messick2017analysis}. These updates were aimed at
decreasing the latency, increasing the sensitivity, and
expanding the parameter space of the pipeline.

\section{Outline of Pipeline Methods} \label{sec:methods}

The GstLAL pipeline makes use of matched filtering~\cite{waistein1970extraction,
thorne1987gravitational, Sathyaprakash:1991mt, Cutler:1992tc,
finn1992detection, finn1993observing, Dhurandhar:1992mw,
Balasubramanian:1995bm, Flanagan:1997sx}, which is a method to extract signals
from noisy data by cross-correlating the detector output with a predicted
waveform signal. In case of gravitational waves from compact binary mergers,
the predicted waveforms come from models that use the Post-Newtonian
approach~\cite{blanchet1995gravitational, blanchet1996gravitational,
blanchet2002gravitational, blanchet2004gravitational,
blanchet2006gravitational}, and the effective-one-body (EOB)
formalism~\cite{buonanno1999effective, PhysRevD.62.064015} to model the
inspiral phase of the waveform. These models are then hybridized with the
results from black hole perturbation theory and numerical relativity to obtain
the full inspiral-merger-ringdown waveforms~\cite{PhysRevLett.106.241101,
PhysRevD.82.064016, PhysRevD.93.044006, PhysRevD.93.044007, PhysRevD.86.024011,
PhysRevD.89.061502, PhysRevD.95.044028}. 

The compact binary coalescence waveforms depend on a number of parameters, some
that are intrinsic to the source, such as the masses and spins of the binary
components, and some that are extrinsic, such as the distance, inclination
angle, etc. which are related to the position of the source with respect to the
observer. When performing matched filtering, our goal is to maximize the
matched-filter output over all these parameters. The extrinsic parameters
enter in the waveform only as an overall amplitude and an overall phase.\footnote{This is true for
systems in which spins of the component objects are aligned or anti-aligned to
their orbital angular momenta. However, when the component spins are misaligned
with the orbital angular momenta, the inclination and polarization angles
become time dependent and lead to a non-trivial amplitude and phase modulation of the
observed signals.} The maximization over the overall coalescence phase in the detector frame is done
analytically, and that over the time of coalescence and all the instrinsic
parameters is done by brute force. For maximizing over the intrinsic
parameters, we create a template bank~\cite{owen1996search, owen1999matched,
harry2009stochastic, ajith2014effectual} containing a discrete set of waveforms
spanning the intrinsic parameter space of our search. These discrete points in
the search parameter space are chosen such that the mismatch between any signal
and the best matching template from the template bank (arising from the
discrete nature of our template bank) is less than a predecided tolerance set as 3\% in O2. In the
template bank for O2, we consider the spins of the objects to be aligned to the total
angular momentum of the system~\cite{privitera2014improving}, although efforts
are ongoing to also include systems which have component spins that are not
aligned to the orbital angular momentum~\cite{schmidt2012towards,
hannam2013twist, PhysRevD.89.084006, Harry:2016ijz}. The template bank that was
used in O2 is described in more detail in~\ref{ss:bank}. Validation of the template bank is discussed in~\cite{2018arXiv181205121M}.

The output of the matched-filter is the signal-to-noise ratio (SNR), the inner
product of the whitened data with the whitened template. In the GstLAL pipeline, it is
calculated in the time-domain~\cite{messick2017analysis}:

\begin{align} x_i(t) &= \int_{-\infty}^\infty \mathrm{d} \tau \,
\hat{h}_i(\tau) \hat{d}(\tau+t), \label{eq:gstlaltimemf} \end{align}

where

\begin{align} \hat{d} (\tau) &= \int_{-\infty}^\infty \mathrm{d}f
\frac{\widetilde{d}(f) }{\sqrt{S_n(|f|)/2}} e^{2 \pi i f \tau} \end{align}

is the whitened data and the whitened template $\hat{h}_i(\tau)$ is defined
similarly. The subscript, $i$, runs over each set of template parameters in our
template bank. $S_n(f)$ is the single-sided noise power spectral density (PSD).
The whitener used in the pipeline is described in~\ref{ss:whitener}. The
whitened data undergoes data quality checks and conditioning as described
in~\ref{ss:autoveto}. For each set of parameter values in the template bank, we
have two real waveforms - one corresponding to the ‘+’ polarization, and the
other corresponding to a `quadrature phase-shifted +' waveform, which is equal
to the ‘$\times$’ polarised waveform barring an overall amplitude factor.

We construct a complex signal-to-noise ratio (SNR) time series, real part of
which is the SNR time series from the `+’ polarized template ($x_i(t)$), and
the complex part is the SNR time series from the `quadrature-phase shifted +'
polarized template ($y_i(t)$). We maximize over the unknown time and phase by
maximizing over the absolute value of the complex SNR time series over time,

\begin{align} \rho &=\max_{t}|x_i(t)+iy_i(t)|. \end{align}
%
The GstLAL pipeline makes use of the LLOID~\cite{cannon2012toward, messick2017analysis} algorithm, which combines singular valued
decomposition with near-critical sampling to construct a reduced set of
orthonormal filters with far fewer samples.

Detector data often contain glitches~\cite{abbott2016detchar} (see also~\ref{ss:autoveto}), which can
produce high peaks in the SNR time series. SNR is not
sufficient to distinguish noise from transient signals in presence of
non-Gaussian data. Therefore, in addition to recording the peaks in the SNR
time series, the pipeline performs a signal consistency check whenever it
records an SNR above a certain threshold. This is done by determining how
similar the SNR time series of the data around the peak value is to the SNR
time series expected from a real signal. The SNR time series is predicted by
calculating the auto-correlation between the complex template waveform and
itself, and scaling it by the peak complex SNR. This predicted SNR time series
is equal to the SNR time series under the assumption that the signal matches
the template waveform exactly in absence of any detector noise.  This signal
consistency test value, $\xi^2$, is computed by integrating the amplitude
squared of the difference between the complex SNR time series and the predicted
SNR time series over a $\delta t$ time window around the peak, and normalizing
it appropriately~\cite{messick2017analysis}, \begin{equation} \xi^2 =
\frac{\int_{-\delta t}^{\delta t} dt \abs{z(t) - z(0)R(t)}^2}{\int_{-\delta
t}^{\delta t} dt (2 - 2\abs{R(t)}^2)}.  \end{equation}

Here $z(t)$ is the complex SNR time series, $z(0)$ its peak, and $R(t)$ is the
auto-correlation series.  Whenever the GstLAL pipeline records a peak in the SNR
time-series that is greater than a preset threshold\footnote{This SNR threshold was set to 4 for the Hanford and Livingston detectors in O2. For the online analysis, the minimum trigger SNR in Virgo was not determined
by an explicit threshold, but instead by a restriction to record at most 1 trigger per second in a given template. For the offline analysis, the minimum SNR threshold was set to 3.5 for Virgo.}, it records the SNR and $\xi^2$,
the masses and the spins of the template that returned those values upon
matched filtering, and the phase and the time of
coalescence. Together these quantities form a `trigger'. These triggers are
divided into smaller bins based on the spins and masses of the templates.
Triggers from different detectors corresponding to the same template that are coincident in time within a
time window which takes into account the maximum light travel time between detectors and statistical fluctuations in the measured event time due to detector noise are called coincident triggers. In O2, a time window of 5 ms plus light travel time between the detectors was used.



The set of triggers that did not participate in a coincidence when more than
one detector was operating are used to characterize noise for their respective
(mass-spin) bin. The GstLAL pipeline builds a signal and a noise model as a function
of the SNR, $\xi^2$, the detector(s) that participated in making the set of
trigger(s), and the horizon distances of all the detectors at the time of the
event.  The signal model is constructed by assuming that the signals follow
their expected distribution in Gaussian noise~\cite{cannon2013method}.  The
expectation value of SNR can be obtained by assuming unform in volume
distribution of sources, and the distribution of $\xi^2$ is obtained by
assuming a maximum of 10\% loss in SNR due to waveform
mismatch~\cite{cannon2013method, allen2005chi}. The triggers are then assigned
a log likelihood-ratio which is the pipeline's detection statistic based on the
noise and the signal model for that particular bin. A more detailed description
of the likelihood-ratio statistic can be found in~\ref{ss:lr}. A  Monte Carlo
sampler is used to draw from these background probability density functions
formed from the single detector triggers to construct a mapping from the log
likelihood-ratio ($\mathcal{L}$) to the false-alarm-probability, 
$FAP(\mathcal{L})$, which is the probability of having at least one event with
a log likelihood-ratio greater than or equal to $\mathcal{L}$ under the noise
hypothesis~\cite{cannon2013method, cannon2015likelihood}.

For a detailed description of these methods we refer the reader
to~\cite{messick2017analysis, cannon2010singular, cannon2012toward,
cannon2015likelihood}. In the following subsections we describe the
developments in these methods used by the GstLAL pipeline since the configuration used
during Advanced LIGO's first observing run~\cite{messick2017analysis}.

\section{Latest Developments} \label{sec:ld}

	\subsection{Template bank decomposition} \label{ss:bank}

As described in ~\ref{sec:methods}, template banks are discrete sets of
waveforms that ensure that the SNR loss due to signal and template waveform
mismatch will not be greater than a pre-specified threshold.

\begin{figure}[hbt!]
\includegraphics[width=0.5\textwidth]{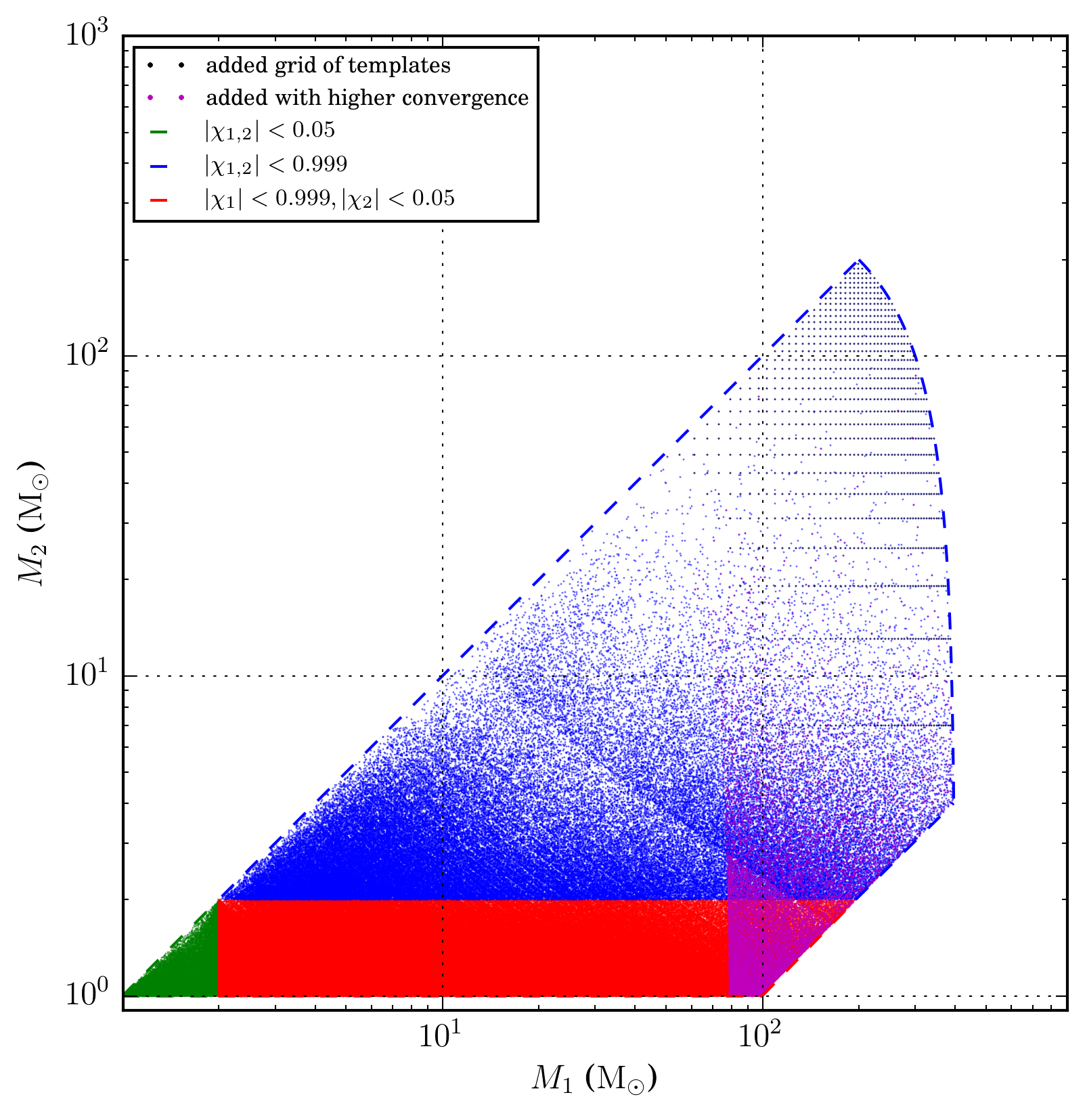}
\caption{\label{bank.png}A visual representation of the bank used in
the second observing run in the component mass space. Each point here represents a template in the bank. All regions shown in the figure are discrete points, they continuous regions being the highly dense regions. The green points are the binary neutron star templates, blue are the binary black hole templates, and red are the neutron star-black hole templates. Magenta and black points represent additional templates that were added to aid in the background estimation for the scarcely populated, high-mass region of the template bank.}
\end{figure}

For Advanced LIGO\textsc{\char13}s second observing run, the parameter space of
the template bank was increased from a total mass of $2M_\odot-100M_\odot$ to
$2M_\odot-400M_\odot$. Neutron stars were assumed to have masses less than
$2M_\odot$ with dimensionless spin parameters in the range (-0.05, 0.05). Black holes were assumed to
have masses greater than $2M_\odot$ with dimensionless spins in the range (-0.999, 0.999).
Individual masses of the systems lie in the range $1-399M_\odot$ with mass
ratios in the range $1-97.989$. The templates were placed in two stages, below a
total mass of $4M_\odot$ inspiral-only templates of Post Newtonian
approximation~\cite{buonanno2009comparison} were layed down first by using a geometric
technique~\cite{brown2012detecting} and then using these templates as a seed for the stoachastic algorithm, while above a total mass of $4M_\odot$ 
full inspiral-merger-ringdown templates of effective-one-body approximation
hybridized with numerical relativity and black hole perturbation
theory~\cite{PhysRevD.95.044028} were placed using stochastic
methods~\cite{ajith2014effectual}. The original templates that were placed are
shown in green,
red, and blue in Fig.~\ref{bank.png}. The colors denote the range of
spins covered by the template bank for those masses.

For the purpose of background estimation in the GstLAL pipeline, we first divide the
entire template bank into several sub banks, known as $\bar{\theta}$ bins,
 that contain ``similar" waveforms. This division is done 
based on their intrinsic parameters. We assume that the templates that belong in the same sub bank respond similarly to noise and noise
statistics are collected for all templates in a bin as a whole. The sub banks are 
typically formed by combining a certain number of “split banks", which are
formed for performing the SVD~\cite{messick2017analysis}. In O2, we originally grouped together 2 split banks containing
500 templates each to form the $\bar{\theta}$ bins. We want the binning method to, a) group
together similar templates to use the LLOID method~\cite{cannon2012toward, messick2017analysis}  for
computationally-efficient time-domain searches, and b) group together templates
with similar noise backgrounds for appropriate FAR estimates. Prior to O2,
the pipeline used two composite parameters, which are a combination of
the four instrinsic parameters, in order to group the waveforms into split banks
- the chirp mass and the effective spin. The chirp mass and the effective spin
  parameter are the leading order terms that describe the phase evolution of
the inspiral part of the waveform according to the Post Newtonian expansion.

The chirp mass $\Mc$ is defined as: \begin{align} \Mc &= \frac{(m_1
m_2)^{3/5}}{(m_1 + m_2)^{1/5}}.  \end{align}

The effective spin parameter is given by: \begin{align} \chi_\mathrm{eff}
&\equiv \frac{m_1 \chi_1 + m_2 \chi_2}{m_1 + m_2}, \end{align}

where $m_1$, $m_2$ are the masses and $\chi_i = \vec{S}_i \cdot \hat{L}/m_i^2$
are the dimensionless spin parameters of each component in the binary.
$\vec{S}_i$ are the component-spin vectors and $\hat{L}$ is the orbital angular
momentum unit vector.

While analyzing the data from O2, it was found that for some analyses,
there was an excess in the closed box result compared to the background
predicted by the GstLAL pipeline. The closed box result is prepared by looking at time-shifted coincidences between the two detectors, and is expected to closely
follow the background model when the pipeline is only treating coincident
events as candidates. After investigations, it was
found that the template bank, together with the grouping scheme of templates
that was used in the first observing run of the Advanced LIGO
detectors~\cite{messick2017analysis} for collecting noise statistics was
inadequate to construct an accurate background model for the high
mass region that was added to the parameter space for O2.
As a result, a different grouping scheme was introduced in O2, and the
template bank was also modified. In
Fig.~\ref{bank.png} we see that the density of the templates decrease
with the increase in the masses of the binaries. This is a typical feature seen
in all template banks. The waveform of a system with smaller masses is longer
in the frequency band Advanced LIGO and Virgo detectors are most sensitive in.
Therefore even a small change in masses can lead to a big mismatch between two
waveforms in this region since there are more cycles in band for which the
match has to be performed. This means that we need more waveforms to cover the
lower mass region of the parameter space.

\begin{figure}[hbt!]
\includegraphics[width=0.5\textwidth]{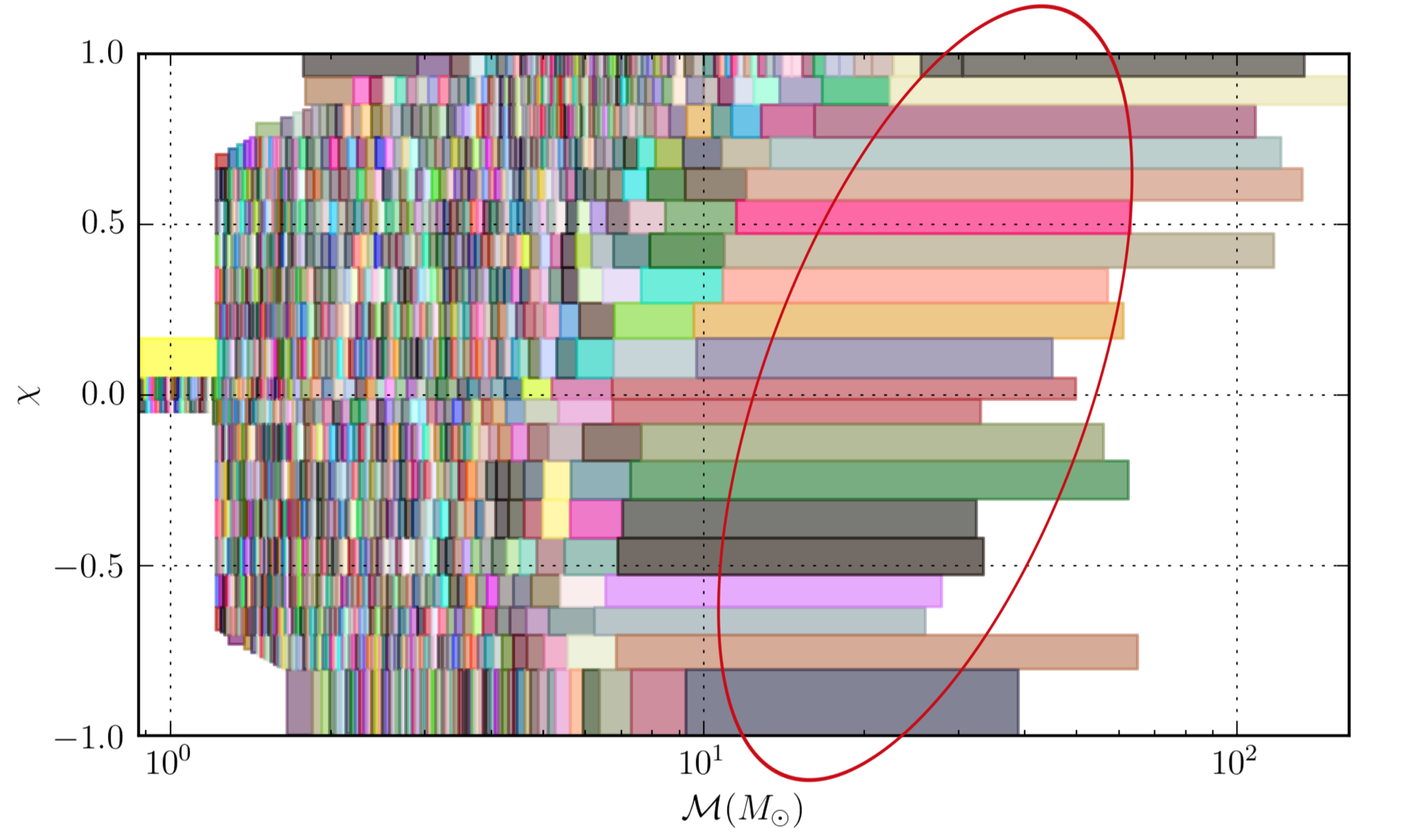}
\caption{\label{oldsvd}A visual representation of the background bins according to the old binning scheme of the templates in $\Mc$ - $\chi_\mathrm{eff}$ space. The bins circled were identified as the ones that contained the templates whose noise properties were not being described by the noise properties of the bins.}
\end{figure}

The extremely low density of the templates in the high mass region of the bank,
which was added for O2 caused  templates vastly
different in $\Mc$, but similar in $\chi_\mathrm{eff}$ to be grouped together in the low template
density region,  since the number
of templates in each bin was required to be the same. This in turn led to sub-optimal estimation of noise properties
for these bins, which led to inaccurate significance estimation of noise
events. In Fig. ~\ref{oldsvd}, these are the highest $\Mc$ groups in each $\chi_\mathrm{eff}$ bin.

\begin{figure}[hbt!]
\includegraphics[width=0.5\textwidth, left]{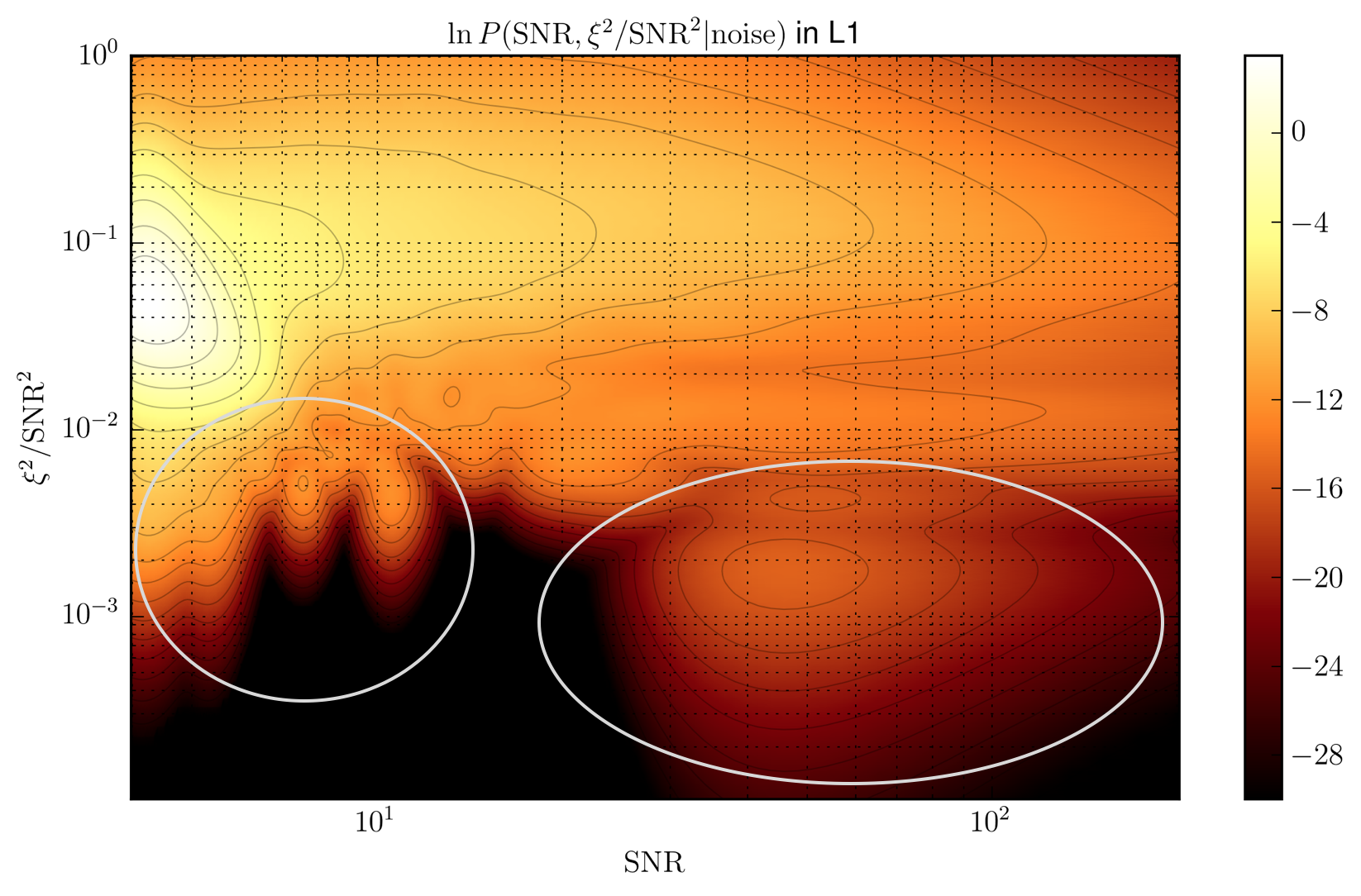}
\caption{\label{badbin}SNR--$\xi^2$ PDF for one of the ``bad" bins. We draw
the reader's attention to some of the island features inside the white circles.
These occur when the pipeline doesn't have enough data about the behavior of
some of the templates in the bin, because respond to noise differently from the
rest of the templates in the bin.}
\end{figure}

We can also see this in the background SNR-$\xi^2$ PDFs for these ``bad" bins that the pipeline uses to assign the likelihood-ratio statistic, see Fig. ~\ref{badbin}. The features marked in white in the PDF happen when there are template present in a bin whose properties are not well represented by the rest of the templates in the bin.

To solve this issue described above, the grouping scheme was changed for the high
mass templates with total mass $>$ 80$M_\odot$. Instead of using $\Mc$ and
$\chi_\mathrm{eff}$, we now use template duration in the LIGO sensitivity band for
grouping these high mass templates. $\Mc$ and $\chi_\mathrm{eff}$ are the
leading order parameters that describe the inspiral part of the waveforms, but
not the merger and ringdown. For high mass systems, a significant amount of power
in the LIGO band comes from merger and ringdown phases of the coalescence,
therefore $\Mc$ and $\chi_\mathrm{eff}$ are no longer the best parameters to group
the templates. The number of templates in these high-mass bins was
also reduced to 200, 400, or 800, instead of 1000 everywhere else, to account for the sparse density of templates in this region.
Hence only those templates that have the similar noise properties are now grouped together. 15,665 extra templates were also added to the
bank above a total mass of 80, by increasing the convergence threshold from 0.97 to 0.98 and by adding a grid of templates. This was done to increase the template density in that region.
This brought the total number of templates to 677000. The templates that were added by increasing the convergence threshold of the stochastic placement process are shown in  magenta and those that were added as a grid are shown in black in Fig. ~\ref{bank.png}. Refer~\cite{2018arXiv181205121M} for a detailed description of the construction of the bank.


	\subsection{Zero Latency Whitener} \label{ss:whitener}

During the first observing run of Advanced LIGO, the pipeline detected a binary black hole merger event, GW151226, with a latency of ~70 seconds~\cite{abbott2016gw151226}, which means that the signal was detected a mere 70 seconds after it arrived on Earth. This was the first time a gravitational-wave event had been identified by a matched-filtering pipeline in low-latency and was a huge success for the pipeline, which aims to send prompt alerts to electromagnetic observatories in order to facilitate studies correlating astrophysical phenomena. But from the recent joint gravitational-wave and gamma-ray detection of GW170817, we now know that the time delay between gravitational-wave emission and the onset of the following SGRB is approximately 2 seconds, motivating achieving alert latencies below 2 seconds in order to capture the earliest associated electromagnetic phenomena in other bands.

The original whitening filter employed in the GstLAL pipeline contributes to one of
the bottleneck processes in the pipeline's latency, adding up to 32
seconds to the total latency time~\cite{messick2017analysis}. The zero latency whitener was introduced in spirit of reducing the latency of the pipeline~\cite{2017arXiv170804125T}.

As described in ~\ref{sec:methods}, a matched filter is the inner product between a template
waveform and the strain data. However, since the strain data are strongly
colored by the frequency-dependent noise of the detector, we `whiten'
the inner product by weighting both the data and the template waveform by a factor of $1/\sqrt{S_n(f)}$ each, where $S_n(f)$
is the single-sided power spectral density of the detector noise. The method of a
frequency-domain whitening filter described in~\cite{messick2017analysis} has discrete Fourier
transforms and window functions applied to 32-second blocks of
input data, with the PSD being updated every 16 seconds. Since a
32-second block is processed every 16 seconds, this filter has a
latency between 16 to 32 seconds, which is more than half of the
pipeline's total latency.

\begin{figure}[hbt!]
\includegraphics[width=0.5\textwidth]{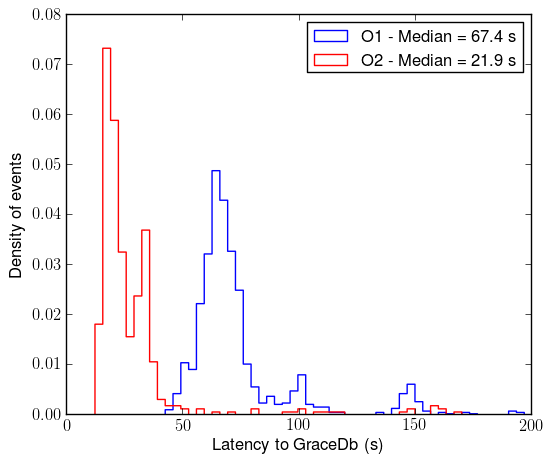}
\caption{\label{latency}Comparison of the latency of detections by the
\texttt{GstLAL} online pipeline in O1 (blue) vs. in O2 (red). We can see that the latency of
the pipeline has been reduced by $\approx$ 40 s. The application of the
zero-latency whitener played a significant contribution in decreasing the
latency of the pipeline.}
\end{figure}

To reduce the latency, we need to whiten in the time-domain. To this end, a Finite Impulse Response (FIR) filter-based algorithm to the frequency-domain whitening is introduced. The square root of inverse PSD of a given strain data is used to construct the FIR of a linear phase filter. This filter still requires 16 seconds of data from the future for its evaluation~\cite{PhysRevD.97.103009}. It is not possible to further reduce the latency of this filter without changing the whitening transformation. Therefore an approximation to the original filter is introduced~\cite{PhysRevD.97.103009}, which derives a minimum-phase approximation of the desired filter~\cite{840000}. The matched filter output is insensitive to the phase response of the whitening filter. The minimum-phase whitening filter accurately approximates the amplitude response of the FIR-based filter, introducing most errors only in the phase response. This filter does not use any information from future samples for its evaluation, and is therefore called a ``zero-latency whitening filter". A new windowing process has also been implemented for the new whitening method. The PSD transition is now allowed to occur continuously, and the resulting filter is a linear combination
of the newest and next newest filters during their transition. This
is recursively applied to the zero-latency algorithm when a new
whitening filter becomes available. It is shown in~\cite{PhysRevD.97.103009} that the causal FIR filter approximation successfully whitens the data, producing zero mean, unit variance, white Gaussian noise. Matched-filter outputs (SNR and $\xi^2$, ~\ref{sec:methods}) produced using the zero-latency whitener are compared to those produced using the frequency-domain whitener for both noise and and simulated signal triggers, and we see a good agreement for both. The time-domain zero-latency filter was used by the GstLAL low-latency analysis of O2, and was a significant contributor in bringing down the latency of the pipeline as compared to O1~\ref{latency}`. The details of the filter, and the consistency checks with the old filter are described in~\cite{PhysRevD.97.103009}.

	\subsection{Data Conditioning} \label{ss:autoveto}


The output of the matched filter is the SNR (~\ref{sec:methods}), which is the optimal detection statistic under the assumption that noise is stationary and Gaussian.
For offline analyses, where the GstLAL pipeline processes archival gravitational-wave data, we use data quality vetoes to flag poor data~\cite{abbott2016detchar}. However, such information is not available for online analyses. Gating on the whitened strain data, whitened $h(t)$, is one of the techniques adopted by the pipeline to eliminate short transient instrumental noise fluctuations. These fluctuations,
also known as glitches, can cause unreasonably high values of SNRs in the data, mimicking gravitational-wave signals, and causing false alarm triggers.

In presence of glitches, the whitened $h(t)$ may have values higher than the expected values from the coalescences of binary systems that the pipeline is aiming to detect. By construction, whitened $h(t)$ should have a unit variance.
Whenever whitened $h(t)$ is momentarily greater than a threshold value, set as some multiple of the standard deviation $\sigma$ of $h(t)$, we gate that piece of data by setting the samples around the peak with a window of 0.25 s on each side to zero~\cite{messick2017analysis}. 

The amplitude of a signal increases with the chirp mass $\Mc$ (~\ref{ss:bank}) of a binary system. Therefore we set the threshold based on the highest masses we are sensitive to, such that it is higher than the whitened strain amplitude we expect from such systems. We want the threshold to be such that it removes the maximum number of glitches from our data without gating out real signals. During the first observing run, the threshold value was set to 50$\sigma$. At lower threshold values, it was seen that the pipeline started to gate some of the high mass simulated BBH signals that were injected in the data.

\begin{figure*}[t]
\includegraphics[width=1.0\textwidth]{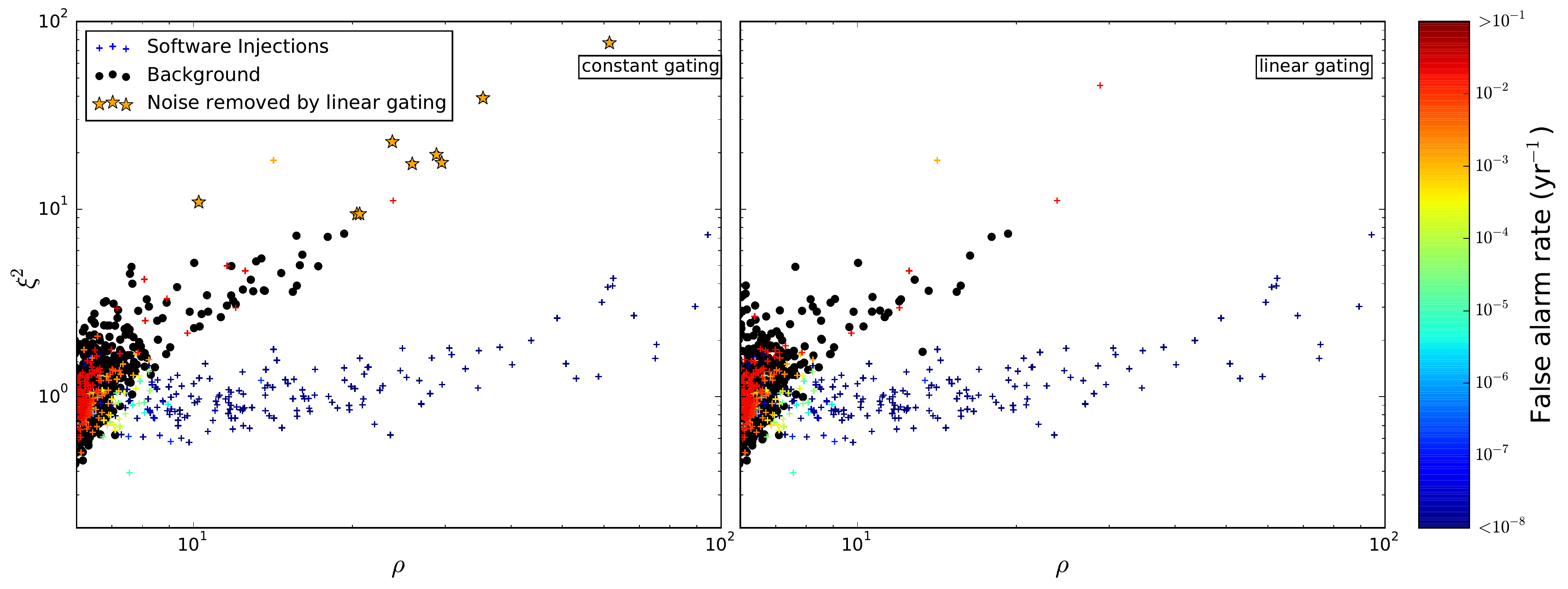}
\caption{\label{fig:ht_gating_comparison}Comparison results between a run with constant gating scheme (left) and linear gating scheme (right). The yellow stars highlight the additional background glitches present in the constant gating run that have been removed in the run with the linear gating scheme applied.}
\end{figure*}

As described in ~\ref{ss:bank}, in the second observing run, the parameter space of our search was increased from a maximum total mass of 100$\msun$ to 400$\msun$. In order to avoid gating the highest mass signals, we would have to increase our value of the gate threshold. This would cause an increase in the number of glitches that pass through without being gated, and therefore an increase in the number of false alarm triggers. Hence a linear gating scheme, in which the  gate threshold value is a linear function of chirp mass instead of a constant for all masses, was introduced.
The pipeline computes an appropriate threshold value according to the linear scale provided by the user and the highest chirp mass template in a sub-bank. Fig.~\ref{fig:ht_gating_comparison} shows that linear gating scheme helps in removing more glitches as compared to the constant gating scheme, while still recovering all the simulated signals injected in the data.

	\subsection{Likelihood-Ratio Statistic} \label{ss:lr}

Once all the candidate events are identified by the pipeline, each of them is assigned a
likelihood-ratio statistic in order to determine its significance. The likelihood
ratio as defined in~\cite{cannon2015likelihood,messick2017analysis} was used in
the analysis of data from the first observing run of Advanced LIGO during which only the Hanford and Livingston detectors were
operating. It is the ratio of probability of certain observables given the signal hypothesis versus the noise hypothesis. Included observables were terms to account for the set of detectors involved in
the coincidence \{H1, L1\}, the detector horizon distances (for a 1.4$\msun$ - 1.4$\msun$ system) at the time of
coincidence $\{D_\mathrm{H1}, D_\mathrm{L1}\}$, the SNRs for each trigger
$\{\rho_\mathrm{H1}, \rho_\mathrm{L1}\}$, and the $\xi^2$-signal-based-veto
values for each trigger $\{\xi^2_\mathrm{H1}, \xi^2_\mathrm{L1}\}$. For Advanced LIGO\textquotesingle s second observing run,
the likelihood-ratio statistic was modified to allow the pipeline to rank single detector
events when only one detector is operational and to include additional
parameters when both advanced LIGO facilities are operational. The inclusion of
additional parameters was done by two methods during O2. The first method which
we will describe shortly in ~\ref{ss:lik} was used to analyze the data in the
beginning of O2~\cite{Abbott:2017vtc, Abbott2017b, PhysRevLett.119.141101, PhysRevLett.119.161101}, and the second method described in ~\cite{Hanna:2019ezx} was
used to analyze final calibrated version of O1 and
O2~\cite{LIGOScientific:2018mvr}. The binning for SNR and $\xi^2$ histograms,
used to compute the $(\rho, \xi^2)$ PDFs, was also changed O2. The $\tan^{-1}
\ln$ binning described in~\cite{cannon2015likelihood} is still used, to collect $\textrm{SNR}-\xi^2$ but now
$x_\mathrm{lo}=2.6$, $x_\mathrm{hi}=26$, and $n = 300$; the values used in O1
were $x_\mathrm{lo}=3.6$, $x_\mathrm{hi}=70$, and $n = 600$. The binning was changed so that the shape of the density kernel estimation matches the natural shape of the $\textrm{SNR}-\xi^2$ PDFs it is being used to estimate.

		\subsubsection{Single Detector Events} \label{sss:sdet}

In Advanced LIGO\textquotesingle s first observing run, the pipeline could only identify gravitational-wave events when both the detectors were operating. Gravitational-wave candidates were formed by demanding coincidence between the two detectors. This meant that we were blind to signals occuring during single-detector time (defined as time when only one detector is operational). For O2, the online analysis also looked at the non-coincident candidates, and these were also assigned a log likelihood ratio statistic.

Non-coincident triggers found during single-detector time (defined as time when only one detector is operational) are now excluded from informing the background model, since these could potentially be loud signals that were found as non-coincident triggers in absence of data from multiple detectors. Since we use an SNR threshold of 4 for triggers, there are too many non-coincident triggers to write to disk. Therefore the non-coincident triggers are first assigned a preliminary log likelihood-ratio, and only those that have log $\mathcal{L}_{prelim}>2$ are considered as gravitational-wave candidates. The likelihood-ratio defined in~\cite{cannon2015likelihood, messick2017analysis} is still valid
for single detector events,
\begin{widetext}
\begin{align}
        \likehood \left( \{D_\text{IFOnet}\}, \{\text{IFO}\}, \rho_\text{IFO}, \xi^2_\text{IFO}, \params \right) &= \likehood \left( \{D_\text{IFOnet}\}, \{\text{IFO}\}, \rho_\text{IFO}, \xi^2_\text{IFO}  \mid  \params \right) \likehood \left( \params \right) \notag \\
        &= \frac{P \left( \{D_\text{IFOnet}\}, \{\text{IFO}\}, \rho_\text{IFO}, \xi^2_\text{IFO} \mid  \params, \sh \right)}{P \left( \{D_\text{IFOnet}\}, \{\text{IFO}\}, \rho_\text{IFO}, \xi^2_\text{IFO} \mid  \params, \nh \right)} \likehood \left( \params \right) \label{eq: likelihood definition}.
\end{align}
\end{widetext}

Here, $\{D_\text{IFOnet}\}$ is the set of horizon distances
for all instruments in the network at the time the event
is  observed, $\{\text{IFO}\}$ is the detector that produced the non-coincident trigger, $\rho_\text{IFO}$, and $\xi^2_\text{IFO}$ are the SNR and $\xi^2$ values of the trigger. The numerator and denominator of the fraction in Eq.~\ref{eq: likelihood definition} are factored in the same way as described in~\cite{cannon2015likelihood, messick2017analysis}. In particular the factorization leads to a form,
\begin{widetext}
\begin{align}
 \likehood \left( \{D_\text{IFOnet}\}, \{\text{IFO}\}, \rho_\text{IFO}, \xi^2_\text{IFO}, \params \right) &=  \dotsm\times\frac{P \left( \{\text{IFO}\} \mid  \{D_\text{IFOnet}\},\params, \sh \right)}{P \left(\{\text{IFO}\}\mid  \params, \nh \right)} \times\dotsm .
\end{align}
\end{widetext}

The probability that a signal yields a trigger above-threshold in only one of the detectors depends on the horizon distances of all the detectors operating at the time, and the duty cycles of all the detectors in the network. The probability that noise yields an above-threshold trigger in one of the detectors is computed from the trigger rates, coincidence window size, and the duty cycles of all the detectors in the network. If more than one detector is operating, and the signal is only seen above-threshold in one of the detectors then that becomes a constraint on the SNR distribution. The single-variable SNR distribution is computed using a Monte Carlo generation of samples described in~\cite{cannon2015likelihood}. In the case where only one detector is operating, the SNR distribution reduces to $P(\rho)\propto \rho^{-4}$.

For assigning the false-alarm rate (FAR), in case of coincident triggers, the pipeline calculates the probability of accidental coincidence by drawing events from each of the single-detector background PDFs. This allows us to measure the likelihood-ratio distribution under the noise hypothesis to values of likelihood-ratio higher than we have actually observed in the experiment. But the non-coincident triggers cannot benefit from the boost due to low probability of an accidental coincidence and therefore the FAR cannot be measured less than $1/\text{(time of experiment)}$. 
GW170817 is an example of the success of including single detector candidate events in the pipeline. It was first identified as a single-detector Hanford event, because of the presence of a glitch in Livingston at the time of the event.

		\subsubsection{Inclusion of phase and time delay terms in the likelihood statistic} \label{ss:lik}

Two additional parameters were added to the likelihood-ratio statistic for the
case where both the Hanford (H1) and Livingston (L1) advanced LIGO detectors are 
operational: $\Delta t$, the difference in end times between the H1 and L1
triggers, and $\Delta \phi$, the difference in coalescence phase between the H1
and L1 triggers. We require $\Delta \phi \in [-\pi, \pi]$ and compute the
modulus $\Delta\phi\,(\mathrm{mod}\,2\pi)$ to enforce a cyclic distribution.

This new likelihood-ratio statistic is defined as follows
\begin{widetext}
\begin{equation} \label{eq:lnL}
	\mathcal{L} = \frac{P \left( \left. \{D_\mathrm{H1}, D_\mathrm{L1}\}, \{ \mathrm{H1, L1}\}, \rho_\mathrm{H1}, \rho_\mathrm{L1}, \xi^2_\mathrm{H1}, \xi^2_\mathrm{L1}, \Delta \phi, \Delta t \right | \mathrm{signal} \right)}
	{P \left( \left. \{D_\mathrm{H1}, D_\mathrm{L1}\}, \{ \mathrm{H1, L1}\}, \rho_\mathrm{H1}, \rho_\mathrm{L1}, \xi^2_\mathrm{H1}, \xi^2_\mathrm{L1}, \Delta \phi, \Delta t \right | \mathrm{noise} \right)}.
\end{equation}
\end{widetext}
This statistic only supports ranking coincidences found with the H1L1 network. Future work for the third observing run will add support for the H1L1V1 network with a goal towards a generalized N-detector network statistic~\cite{Hanna:2019ezx}. 
Additionally, we make several assumptions when factoring the dependencies of the probability density functions in Eq.~\ref{eq:lnL}.
We assume that the noise distributions of $\Delta t$ and $\Delta
\phi$ are independent of each other. We assume that the $\xi^2$ statistic is dominated by 
instrumental noise, thus the $\xi^2$ term reduces to its previous form given in~\cite{messick2017analysis}. We expect that the signal (but not the noise) distributions for $\Delta t$
depend on trigger SNRs as well as on detector sensitivities. The SNR ratio from the two detectors for a signal depends on the position of the source with respect to the detectors and the inherent sensitivities of the detectors. $\Delta t $ for a signal depends only on the position of the source and the location of the observatories. Thus we model the $\Delta t$ distributions as a function of a ratio of SNRs normalized by
horizon distances, to factor out the inherent sensitivities of the detectors, so this term only depends on the position of the source with respect to the detectors. Furthermore, we define it such that it is always smaller than 1,
\begin{equation}
\rho_\mathrm{ratio} = \mathrm{min}(\frac{\rho_\mathrm{H1}/D_\mathrm{H1}}{\rho_\mathrm{L1}/D_\mathrm{L1}}, \frac{\rho_\mathrm{L1}/D_\mathrm{L1}}{\rho_\mathrm{H1}/D_\mathrm{H1}})
\end{equation}
On the other hand, we do not consider dependence on
the detector sensitivities when modeling the $\Delta \phi$ distributions. We only consider dependence of $\Delta \phi$  on
$\Delta t$ and network SNR, defined as
\begin{equation}
\rho_\mathrm{network} = \sqrt{\rho^2_1 + \rho^2_2}
\end{equation}

With these assumptions, the factor describing the dependence of the likelihood-ratio in terms of $\rho$, $\Delta t$, and $\Delta \Phi$ can be written as,
\begin{widetext}
\begin{equation} \label{eq:lnLfactor}
         \begin{split}
	\mathcal{L} & \propto \frac{P \left( \left. \Delta t \right| \{ \rho \}, \{\mathrm{Horizon}\}, \mathrm{signal} \right) P \left( \left. \Delta \phi \right| \Delta t, \{ \rho \}, \{\mathrm{Horizon}\}, \mathrm{signal} \right)}{P \left( \left.  \Delta t \right| \mathrm{noise} \right) P \left( \left.  \Delta \phi \right| \mathrm{noise} \right)} \\
	& \approx  \frac{P \left( \left. \Delta t \right| \{ \rho_\mathrm{ratio} \}, \mathrm{signal} \right) P \left( \left. \Delta \phi \right| \Delta t, \{ \rho_\mathrm{network} \}, \mathrm{signal} \right)}{P \left( \left.  \Delta t \right| \mathrm{noise} \right) P \left( \left.  \Delta \phi \right| \mathrm{noise} \right)} \\ 
	\end{split}
	\end{equation}
\end{widetext}

\begin{figure}
  \includegraphics[width=\columnwidth]{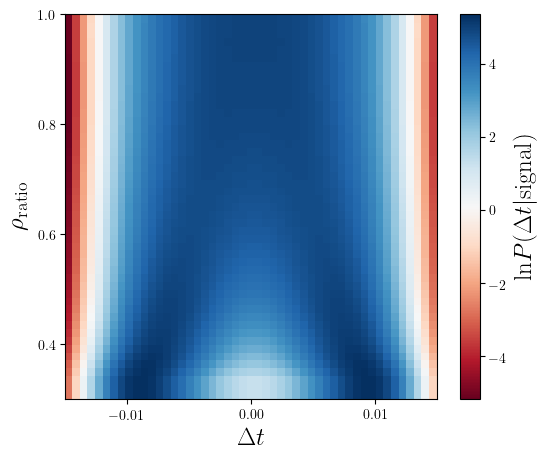}.
  \caption{\label{fig:lnpdt}Distribution of $P \left( \Delta t | \{ \rho_\mathrm{ratio} \}, \mathrm{signal} \right)$}
\end{figure}

\begin{figure}
  \includegraphics[width=\columnwidth]{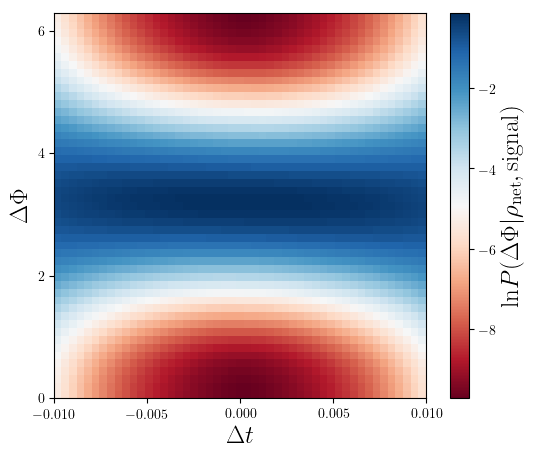}
  \caption{\label{fig:lnpdphi}Distribution of $P \left(  \Delta \phi | \Delta t, \{ \rho_\mathrm{network} \}\right)$ where we have set $\rho_\mathrm{H1}=10$ and  $\rho_\mathrm{L1}=10$.}
\end{figure}

In order to construct distributions for $P \left( \Delta t | \{ \rho_\mathrm{ratio} \}, \mathrm{signal} \right)$ and $P \left(  \Delta \phi | \Delta t, \{ \rho_\mathrm{network} \}, \mathrm{signal} \right)$, we performed an injection (simulated signal injected in the data to test the performance of the pipeline) campaign in the data from the first observing run covering the parameter space of binary neutron star and binary black hole systems. We discarded any injection recovered with $ | \Delta t |  >0.01$ s, assuming that this regime is dominated by accidental coincidences. The $\Delta t$ histograms were smoothed with a gaussian kernel such that they taper to zero above $\abs{\Delta t} = 0.01$ s. Smoothed, normalized $ \Delta t $ histograms were modeled as a function of $\rho_\mathrm{ratio}$ using Chebyshev polynomials. Fig.~\ref{fig:lnpdt} shows the logarithm of the PDF for $P \left( \Delta t | \{ \rho_\mathrm{ratio} \}, \mathrm{signal} \right)$ as defined over a range of $ \Delta t $ and $\rho_\mathrm{ratio}$. Normalized histograms of $\Delta \phi$ were modeled using von Mises distributions which is a continuous probability distribution on a circle since $\Delta \phi$ is cyclic, plus a uniform noise background as a function of  $\Delta t$ and $\rho_\mathrm{network}$. In Fig.~\ref{fig:lnpdphi}, we see the logarithm of the PDF for $P \left( \Delta \phi | \Delta t, \{ \rho_\mathrm{network} = 14\} \right)$ as defined over a range of $ \Delta t $ and $\Delta \phi$.

In order to construct the distributions for $P \left( \Delta t | \mathrm{noise} \right)$ and  $P \left(  \Delta \phi | \mathrm{noise} \right)$, we simply assume $\Delta t$ and $\Delta \phi$ are uniformly distributed for noise triggers that form false coincidences.

        	\subsubsection{Computing joint SNR PDF for different horizon distance ratios} \label{ss:hd}

The horizon distance $D_h$ is the effective distance at which a binary system
is observed with
a nominal SNR of 8~\cite{allen2012findchirp}. This means
that horizon distance is a measure of a detector's sensitivity to a particular system. The horizon distances included in the ranking statistic of the pipeline are computed for a $1.4M_{\odot}-1.4M_{\odot}$ binary neutron star system, and their
fluctuations reflect fluctuations in the
noise spectrum.

One of the factors in the numerator of the likelihood-ratio ranking statistic is
the joint SNR PDF given a set of horizon distances of all the detectors at the time of the event, and the set of detectors that observed the event with an SNR above the threshold.~\cite{cannon2015likelihood}. For candidates that arise from
genuine signals, the joint SNR PDF depends only on the ratios of the
horizon distances. In the first observing run, the pipeline was limited to two detectors
and assumed a fixed joint SNR PDF, corresponding to equal horizon
distances. This assumption implies that the fractional change in
horizon distance is approximately independent of the mass of the
system being observed, which is not valid because changes to the noise
spectrum do not rescale equally over the entire frequency range. In
the second observing run, this assumption was relaxed by pre-computing joint SNR PDFs for a
collection of discrete horizon distance ratios.

	\subsection{Software injections}

The sensitivity of the pipeline is measured by injecting simulated
gravitational-wave signals into the data and measuring the fraction that is recovered,
as described in~\cite{messick2017analysis}. To reduce the computational
cost of searching for injections, the option to search only a subset of the
template bank was added to the pipeline. The subset is specified by providing a
lower and upper bound on chirp mass (~\ref{ss:bank}). Low-mass injections, where the
gravitational wave is dominated by the inspiral stage, are typically recovered
with more accurate chirp masses than high mass injections. For this reason, the
bounds are computed as a function of injection chirp mass,
$\mathcal{M}_{\mathrm{inj}}$. The function was written based on visual
inspection of chirp mass recovery plots from the searches in the first observing run, and subsequent tests
confirmed a loss of less than 1\% of the total found injections.
\begin{subequations}
	\begin{equation}
		\mathcal{M}_{\mathrm{lower\,bound}} = \begin{cases} 0.65 \mathcal{M}_{\mathrm{inj}}, & \mathcal{M}_{\mathrm{inj}} < 10 \\ 0.5 \mathcal{M}_{\mathrm{inj}}, & 10 \leq \mathcal{M}_{\mathrm{inj}} < 20 \\ 0.5 \mathcal{M}_{\mathrm{inj}}, & \mathcal{M}_{\mathrm{inj}} \geq 20 \end{cases} 
	\end{equation}
	\begin{equation}
		\mathcal{M}_{\mathrm{upper\,bound}} = \begin{cases} 1.35 \mathcal{M}_{\mathrm{inj}}, & \mathcal{M}_{\mathrm{inj}} < 10 \\ 1.5 \mathcal{M}_{\mathrm{inj}}, & 10 \leq \mathcal{M}_{\mathrm{inj}} < 20 \\ 2 \mathcal{M}_{\mathrm{inj}}, & \mathcal{M}_{\mathrm{inj}} \geq 20 \end{cases} 
	\end{equation}
	\label{eq:mchirpbounds}
\end{subequations}

	\subsection{Introducing Virgo} \label{ss:coinc}
The Advanced Virgo~\cite{acernese2015advanced} detector joined the second observing run of the Advanced LIGO detectors on August 1st, 2017. It operated at a lower sensitivity relative to the Advanced LIGO detectors for the observing run reported here. Initially only the online search filtered over the Virgo data stream. Due to this, there was no minimum SNR threshold for a trigger in Virgo. Instead, it was restricted to record at most one trigger per second per template~\cite{LIGOScientific:2019gag}. 

In the first observing run, gravitational-wave candidates were formed by demanding coincidence (both in time and template) between the LIGO Hanford and LIGO Livingston triggers~\cite{messick2017analysis}, [~\ref{sec:methods}]. In the second observing run, this was generalized so that a candidate can be formed by an arbitrary number of detectors. For a network of detectors, we can define different types of coincidences, based on the number of detectors participating in the coincidence. However, due to the lower sensitivity of Virgo in the second observing run and the incompatibility of the likelihood-ratio statistic in the beginning of O2, if Virgo participated in a coincidence with either one of the two detectors, the set of triggers was still considered as a single-detector, non-coincident event. In other words $HV$ and $LV$ doubles were treated as $H$ and $L$ singles respectively. And if Virgo participated in a triple event ($HLV$ triple), it was still treated as an $HL$ double. Nonetheless a network of three detectors improves the sky localization of the source. GW170814 was the first gravitational-wave event that had a significant SNR in Virgo. Including Virgo in the analysis helps in reducing the area of the 90\% credible region from 1160 $\deg^{2}$ when using only the two LIGO detectors to 60 $\deg^{2}$ when  using all three detectors for the binary black hole event GW170814~\cite{PhysRevLett.119.141101}.

For the final offline reanalysis of the O1 and O2 data calibrated with the final versions~\cite{LIGOScientific:2018mvr}, the Advanced Virgo data stream was treated in the same manner the same as the Advanced LIGO detectors and also used to inform the likelihood-ratio statistic. To this end, a new method for calculating the likelihood-ratios was introduced which also uses the information from Virgo. This method is fast, computationally efficient and can also be adapted to addition of new detectors in the network. For the detailed description of this method, we refer the readers to~\cite{Hanna:2019ezx}. In the offline reanalysis, GW170818 was detected as a triple coincident event with an SNR of 4.2 in Virgo, 4.1 in Hanford and 9.7 in Livingston~\cite{LIGOScientific:2018mvr}.

\section{Conclusion} \label{sec:conclusion}

The GstLAL pipeline is a stream-based matched-filtering pipeline that has
detected gravitational waves from several compact binary mergers in near real time
since Advanced LIGO's first observing run. In this work, we have described the
advancements made in the techniques of the pipeline to reduce the latency,
increase the parameter space, analyze single-detector time, incorporate the
Advanced Virgo detector's data stream in the analysis, add new parameters to
the likelihood-ratio statistic, and introduce a template mass-dependent
glitch-excision thresholding method. These methods were successfully deployed
in Advanced LIGO's second and Advanced Virgo's first observing run.

\section{Future Work}

Active development tasks aim to maximize the science from future observations,
including the nature of prompt electromagnetic emissions associated with binary
neutron star and neutron star-black hole mergers. To this end, we are working
on further reducing the latency of the pipeline and providing early warning
alerts for binary neutron stars and neutron star-black hole coalescences 10 s
to a minute before merger, with an approximate sky location so that the
electromagnetic facilities can start the process of setting up their
observations in advance of the merger making it possible to capture the
earliest possible light with narrow field instruments~\cite{cannon2012toward}.
We are also working to add a factor to the likelihood-ratio that takes in as an
input a source population mass model~\cite{FongThesis}, which should make the
pipeline more robust if we believe that these population models are correct.
With an increase in the number of compact binary detections made by the
ground-based interferometers, our population models are expected to
improve~\cite{LIGOScientific:2018jsj}, and including these as a factor in the
likelihood-ratio will in turn aid in an increase in the number of detections.

\section{Acknowledgements} \label{sec:ack}

LIGO was constructed by the California Institute of Technology and
Massachusetts Institute of Technology with funding from the National Science
Foundation (NSF) and operates under cooperative agreement PHY-0757058. We would
like to thank Tito Dal Canton for provinding helpful comments and suggestions.
We would also like to thank Graham Woan for his help in the review of the O2
pipeline. SS was supported in part by the LIGO Laboratory and in part by the
Eberly Research Funds of Penn State, The Pennsylvania State University,
University Park, PA 16802, USA. DM, JC, PB, and DC were supported by the the
NSF grant PHY-1607585. SC was supported by the research programme of the
Netherlands Organisation for Scientific Research (NWO). HF was supported by the
Natural Sciences and Engineering Research Council of Canada (NSERC). CH was
supported in part by the NSF through PHY-1454389.  Funding for this project was
provided by the Charles E.  Kaufman Foundation of The Pittsburgh Foundation.
TGFL was partially supported by a grant from the Research Grants Council of the
Hong Kong (Project No. CUHK 14310816 and CUHK 24304317) and the Direct Grant
for Research from the Research Committee of the Chinese University of Hong
Kong. AG acknowledges support from NSF Grants AST-1716394 and AST-1708146. MW
was supported by NSF grant PHY-1607178. This paper carries LIGO Document Number
LIGO-P1700411.

\appendix \label{sec:appendix}

\bibliography{references}

\begin{acronym}
\acro{GW}{gravitational-wave}
\acro{LSC}{LIGO Scientific Collaboration}
\acro{PSD}{power spectral density}
\acro{SNR}{signal-to-noise ratio}
\acro{SVD}{singular value decomposition}
\acro{aLIGO}{advanced LIGO}
\acro{FAR}{false-alarm rate}
\acro{FAP}{false-alarm probability}
\end{acronym}

\end{document}